\documentclass[10pt,byrevtex,twocolumn,noshowpacs,titlepage,prb]{revtex4}
\usepackage{amssymb}
\usepackage{amsmath}
\usepackage{dcolumn}
\usepackage{bm}
\usepackage{longtable}
\usepackage{graphicx}

\graphicspath{{converted_graphics/}}

\begin{document}

\title{Isotropic superconducting state and high critical currents in Fe$%
_{1+y}$Te$_{1-x}$S$_{x}$ single crystals.}
\author{Rongwei Hu$^{1,2}$, J. B. Warren$^{3}$ and C. Petrovic$^{1}$}
\affiliation{$^{1}$Condensed Matter Physics and Materials Science Department, Brookhaven
National Laboratory, Upton, NY 11973}
\affiliation{$^{2}$Department of Physics, Brown University, Providence, RI 02912}
\affiliation{$^{3}$Instrumentation Division, Brookhaven National Laboratory, Upton, NY
11973}
\date{\today }

\begin{abstract}
We report single crystal growth and a study of superconducting properties in
Fe$_{1+y}$Te$_{1-x}$S$_{x}$. We demonstrate the smallest upper critical
field anisotropy, $\gamma =H_{c2}^{\parallel c}/H_{c2}^{\perp c},$ among all
iron based superconductors, the value of $\gamma $ reaches 1.05 at $%
T=0.65T_{C}$ for Fe$_{1.09}$Te$_{0.89}$S$_{0.11}$, while still maintaining
large values of upper critical field. 
\end{abstract}

\maketitle

The discovery of superconductivity in
quaternary iron based layered superconductor LaFeAsO$_{1-x}$F$_{x}$ with T$%
_{C}$ = 26 K stimulated an intense search for superconductors with higher T$%
_{C}$ in this materials class.\cite{Hosono} Shortly after, the critical
temperatures were raised up to 55 K in materials of \ the ZrCuSiAs structure
type (Ref. 2), and superconductivity had been discovered in Ba$_{1-x}$K$_{x}$%
Fe$_{2}$As and LiFeAs.\cite{Rotter}$^{,}$\cite{Wang}

Superconductivity in the PbO - type FeSe opened another materials space in
the search for iron based superconductors.\cite{Hsu} This was followed by
the discovery of superconductivity in FeTe$_{1-x}$Se$_{x}$ and FeTe$_{1-x}$S$%
_{x}$.\cite{Yeh}$^{,}$\cite{Mizuguchi} Iron chalcogenide superconductors
with simple binary crystal structure share the most prominent
characteristics of iron arsenide compounds: a square planar lattice of Fe
with tetrahedral coordination similar to LaFeAsO or LiFeAs, and Fermi
surface topology.\cite{Zhang} Superconducting T$_{C}$'s up to 37 K were
discovered with the application of hydrostatic pressure.\cite{Margadonna} It
was noted, however, that it would be desirable to have isotropic
superconductors with high $T_{c}$ and ability to carry high critical
currents for power applications.\cite{BES} In this work we report the
synthesis of Fe$_{1+y}$Te$_{1-x}$S$_{x}$ ($x=0-0.14$) superconducting single
crystals. We demonstrate small values of $\gamma =H_{c2}^{\parallel
c}/H_{c2}^{\perp c}$ while still maintaining large values of the upper
critical field and critical currents.

Single crystals of Fe$_{1+y}$Te$_{1-x}$S$_{x}$ were grown from Te-S self flux
using a high temperature flux method.\cite{Canfield}$^{,}$\cite{Fisk}
Elemental Fe, Te and S were sealed in quartz tubes under partial argon
atmosphere. The sealed ampoule was heated to a soaking temperature of
430-450 $^{o}C$ for 24h, followed by a rapid heating to the growth
temperature at 850-900$^{o}C$, and then slowly cooled to 800-840 $^{o}C$.
The excess flux was removed from crystals by decanting. Plate-like crystals
up to $11\times 10\times 2mm^{3}$ can be grown. Elemental analysis and
microstructure was performed using energy dispersive x-ray spectroscopy
(EDS) in an JEOL JSM-6500 scanning electron microscope. The average
stoichiometry was determined by examination of multiple points on the
crystals. Powder X-ray diffraction (XRD) was measured using a Rigaku
Miniflex with Cu K$_{\alpha }$ radiation ($\lambda =1.5418$ \AA ). The unit
cell parameters were obtained by fitting the XRD spectra with the Rietica
software.\cite{Hunter} Flux - free rectangular shaped crystals with the
largest surface orthogonal to c axis of tetragonal structure were selected
for four-probe resistivity measurements. Thin Pt wires were attached to
electrical contacts made with Epotek H20E silver epoxy. Sample dimensions
were measured with an optical microscope Nikon SMZ-800 with 10 $\mu $m
resolution. Magnetization and resistivity measurements were carried out in a
Quantum Design MPMS-5 and a PPMS-9 for temperatures from 1.8 K$\ $to 350 K.

\begin{figure}[tbp]
\includegraphics[scale=0.65]{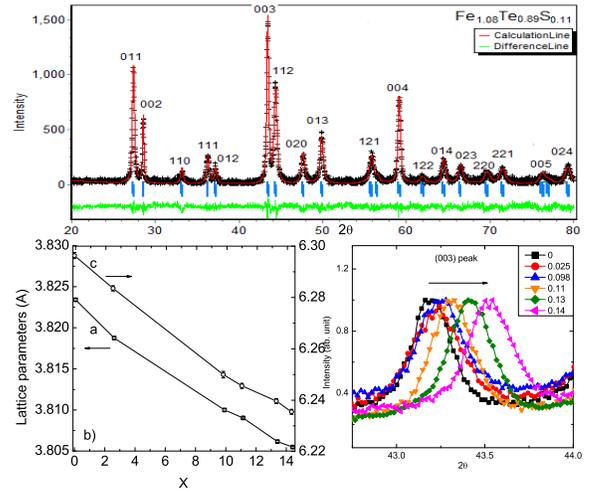} 
\vspace*{-0.5cm}
\caption{Powder X-ray diffraction spectra for $x=0$ and $0.133$ samples.
The lattice parameters vs. S concentration.}
\end{figure}

\begin{figure}[tbp]
\includegraphics[scale=0.65]{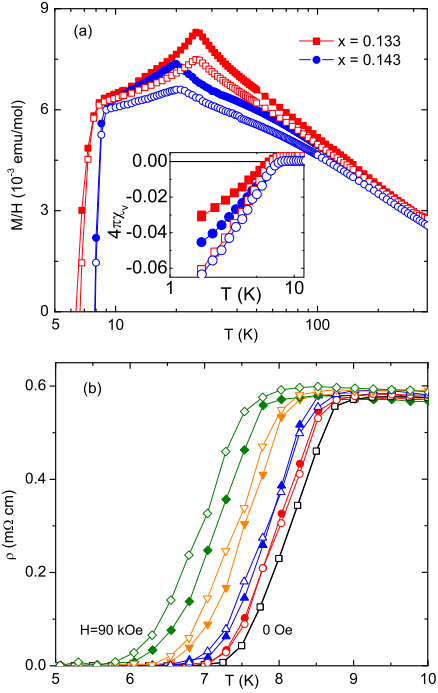} 
\vspace*{-0.5cm}
\caption{a) Temperature dependence of magnetic susceptibility of$~$two
superconducting samples ($x=0.133$ and $0.143$) for $H\perp c$ (solid
symbols) and $H\parallel c$ (open symbols). Inset shows the volume fraction
of the as a function of temperature (1.8-12 K). b) In-plane resistivity for $%
x=0.133$ of two field orientations.}
\end{figure}

\begin{figure}[tbp]
\includegraphics[scale=0.65]{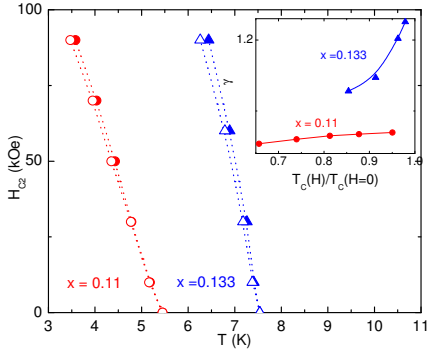} 
\vspace*{-0.5cm}
\caption{The upper critical fields for $x=0.011$ and $0.133.$ Dotted lines are guides to the eye. Inset shows
the anisotropy in the upper critical field $\gamma =H_{c2}^{\parallel
c}/H_{c2}^{\perp c}.$}
\end{figure}

\begin{figure}[tbp]
\includegraphics[scale=0.65]{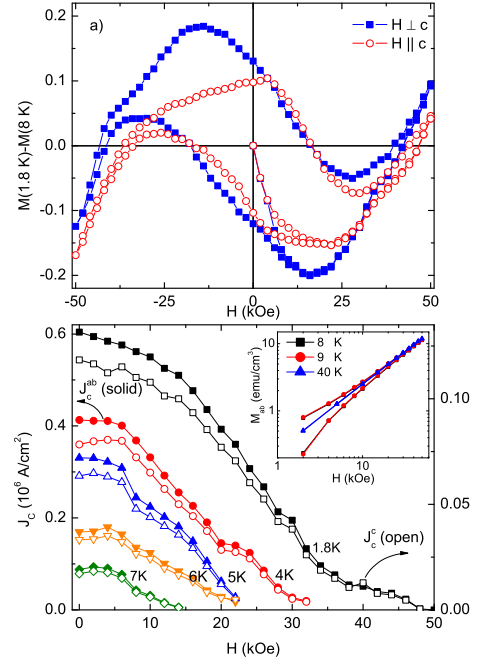} 
\vspace*{-0.2cm}
\caption{a) Magnetization hysteresis loops of $x=0.133$ for $1.8K$ after
ferromagnetic background subtraction for $H//c$ (open symbols) and $H\perp c$
(solid symbols). b) In-plane (to left axis) and interplane (to right axis)
critical currents for $x=0.133.$ Inset shows the magnetization at 8, 9 and
40 K. Only the positive field magnetization is shown on log-log scale and
the virgin curves of the loops at 8 and 9 K are omitted for clarity.}
\end{figure}

The powder X-ray diffraction patterns for all samples investigated can be
indexed in the $P4/nmm$ space group of PbO structure type and small fraction
of Te from the flux. Fig. 1(a) shows refinement of Fe$_{1.09}$Te$_{0.89}$S$%
_{0.11}$. The calculated diffraction pattern shows excellent agreement with
the experiment and high phase purity. The lattice parameters are shown in
Fig. 1(b). Both the a and c axis decrease uniformly with x, conforming with
Vegard's law. Fig. 1(c) shows uniform evolution of $[003]$ peak with sulfur
content, consistent with the decreasing c axis parameter. Thus S doping is
equivalent to a positive chemical pressure and reduces the unit cell volume
by up to 2\%. The compositions from EDS are presented in Table I. The excess
Fe decreases with increasing S content.

\begin{table*}[tbph]
\begin{tabular}{cccccccc}
\hline\hline
$x$ & $y$ & $H_{c2}^{^{\prime }\perp c}(T_{c})$ & $H_{c2}^{\perp c}(0)(T)$ & 
$H_{c2}^{^{\prime }\parallel c}(T_{c})$ & $H_{c2}^{\parallel c}(0)(T)$ & $%
\xi ^{\perp c}(0)(nm)$ & $\xi ^{\parallel c}(0)(nm)$ \\ \hline
$0$ & $0.15(3)$ &  &  &  &  &  &  \\ 
$0.110(6)$ & $0.09(3)$ & $-4.9$ & $19$ & $-4.6$ & $18$ & $4.3$ & $4.3$ \\ 
$0.133(6)$ & $0.08(3)$ & $-10.7$ & $56$ & $-8.4$ & $44$ & $2.4$ & $2.7$ \\ 
\hline\hline
\end{tabular}%
\caption{{\protect\small EDX analysis results , upper critical field at zero
temperature and corresponding coherence length.}}
\end{table*}

Fig. 2 (a) shows the temperature dependence of the magnetic susceptibility
for a magnetic field applied in the ab plane and along c axis. Magnetic
transition is suppressed to 25 K, as was observed in FeTe$_{1-x}$S$_{x\text{ 
}}$polycrystals.\cite{Mizuguchi} Superconductivity is observed for $x\geq
0.11.$ The volume fraction $4\pi \chi _{v}$ reaches $-0.06\sim -0.09$ at 0 K
by linear extrapolation. Considering that the estimated superconducting
volume fraction is only up to 10\%, we speculate that the superconducting
phase may exist in fractional domains associated with S atoms. With more
S-doping, these domains are gradually connected to form a full
superconducting path for electrical transport, as it was observed in CaFe$%
_{1-x}$Co$_{x}$AsF.\cite{Xiao}$^{,}$\cite{Takeshita}

The temperature dependence of resistivity with a magnetic field applied
perpendicular and parallel to c axis is shown in Fig. 2 for $x=0.133.$ The
residual resistivity of the normal state $\rho _{0}=0.58m\Omega cm$ of our
crystals is smaller than that is observed in polycrystalline FeTe$_{1-x}$S$%
_{x}$.\cite{Mizuguchi} It is comparable with residual resistivity observed
in NdFeAsO$_{0.7}$F$_{0.3}$ ($\sim 0.2m\Omega cm$)\cite{Jaroszynski} or (Ba$%
_{0.55}$K$_{0.45}$)Fe$_{2}$As$_{2}$ ($\sim 0.4m\Omega cm$)\cite{Ni} single
crystals. Transition width of our crystals ($\Delta
T_{c}=T_{onset}-T_{zero\rho }=1.8K)$ is smaller than that in Ref. 7 ($\Delta
T_{c}=2K$). The small shift of the transition temperature with magnetic
field indicates a large zero-temperature upper critical field. The upper
critical field $H_{c2}$ is estimated as the field corresponding to the $90\%$
of resistivity drop (Fig.3). An estimate for $H_{c2}(T=0)$ is given by
weak-coupling formula for conventional superconductors in the
Werthamer-Helfand-Hohenberg model (Table I): $H_{c2o}(0)\sim
-0.7H_{c2}^{^{\prime }}(T_{c})T_{c}$.\cite{WHH}\ The superconducting
coherence length $\xi (0K)$ $[\xi ^{2}=\Phi _{0}/2\pi H_{c2}]$ is around 3
nm. The anisotropy $\gamma =H_{c2}^{\parallel c}/H_{c2}^{\perp c}$ decreases
with a temperature decrease approaching a value close to unity. By $%
T_{c}/T_{c}(0)\approx 0.65$ (Fig. 3(inset)), $\gamma =1.05$, for $x=0.11$.
These values indicate that FeTe$_{1-x}$S$_{x}$ is a high field isotropic
superconductor, with $\gamma $ smaller than that in Ref. 19 ($%
\gamma >1.5$ at $0.5T_{C}(H=0))$ or in Ref. 20 ($\gamma \sim 1.3$
at $0.5T_{C}(H=0)).$

To determine the anisotropy of the critical current, we analyze the magnetic
measurements using an extended Bean model.\cite{Bean}$^{,}$\cite{Gyorgy}
Considering a rectangular prism-shaped crystal of dimension $c<a<b$, when a
magnetic field is applied along the crystalline c axis, the in-plane
critical current density $j_{c}^{ab}$ is given by%
\begin{equation*}
j_{c}^{ab}=\frac{20}{a}\frac{\Delta M_{c}}{(1-a/3b)}
\end{equation*}%
in which $\Delta M_{c}$ is the width of the magnetic hysteresis loop for
increasing and decreasing field. When the magnetic field is applied along
the b axis and parallel to the ab plane, both of the in-plane $j_{c}^{ab}$
and the cross-plane $j_{c}^{c}$ are involved in the Bean model. For a
crystal in our measurements with $a=1.245mm,b=1.285mm$ and $\ c=0.732mm$, 
\begin{equation*}
j_{c}^{c}=\frac{c}{3a}\frac{j_{c}^{ab}}{(1-20\Delta M_{b}/cj_{c}^{ab})}
\end{equation*}%
Because of the large volume fraction of the normal and magnetic state, a
magnetic background is superposed on the hysteresis loop. Moreover, as shown
in Fig. 4 (b) inset, the hysteretic magnetisation loop for the sample $%
x=0.133$ sustains above the superconducting transition temperature at 7.5 K
and vanishes above the antiferromagnetic transition at 25 K. It implies a
magnetic structure of FeTe$_{1-x}$S$_{x}$ where a ferromagnetic component
coexists with an antiferromagnetic moment. Density functional calculation on
FeTe by Alaska Subedi et al does indicate that besides the SDW, FeTe is
close to a ferromagnetic instability, similar to LaFeAsO.\cite{Subedi} In
order to estimate the $\Delta M$ only due to flux pinning, we take the
hysteresis loop immediately above superconducting transition at 8 K as the
ferromagnetic background and subtract it from other loops below 7.5 K. The
identical hysteresis loops at 8 and 9 K in the normal state justifies our
rationale to use them as a temperature independent background. Fig. 4(a)
shows hysteresis loops for $H//c$ and $H\perp c$ at 1.8 K after background
removal. The magnetically deduced in-plane and interplane critical current
density are displayed in Fig. 4(b). The ratio of $j_{c}^{ab}/j_{c}^{c}$ is
roughly about 4. The critical current densities for both directions are $%
10^{5}-10^{6}A/cm^{2},$ comparable to MgB$_{2}$, Ba(Fe$_{1-x}$Co$_{x}$)$_{2}$%
As$_{2}$ in the same temperature range\cite{Tanatar}.

In summary, we show that FeTe$_{1-x}$S$_{x}$ are isotropic high field
superconductors with one of the smallest values of $\gamma
=H_{c2}^{\parallel c}/H_{c2}^{\perp c}$ observed so far in iron based
superconducting materials . Moreover, anisotropy in the superconducting
state decreases with increased sulfur content. By utilizing high pressure
synthesis techniques even higher T$_{C}$'s, upper critical fields and
smaller\ $\gamma $ may be simultaneously obtained. Since FeTe$_{1-x}$S$_{x}$
superconductors consist of relatively inexpensive and nontoxic elements they
may represent future materials of choice for high field power applications.

We thank Paul Canfield, Sergey Bud'ko and Myron Strongin for useful
discussions. This work was carried out at the Brookhaven National
Laboratory, which is operated for the U.S. Department of Energy by
Brookhaven Science Associates (DE-Ac02-98CH10886). This work was supported
by the Office of Basic Energy Sciences of the U.S. Department of Energy.

\end{document}